\documentclass[prl,unsortedaddress,superscriptaddress,showpacs,twocolumn,floatfix]{revtex4-1}

\usepackage{graphicx}
\usepackage{amssymb}
\usepackage{amsmath}

\newcommand{\Pmu}{\ensuremath{{\cal P}_{\mu}}}
\newcommand{\Ppimu}{\ensuremath{{\cal P}^{\pi}_{\mu}}}
\newcommand{\scinot}[2]{\ensuremath{#1\!\times\!10^{#2}}}

\begin{document}

\title{New Experimental Constraints for the Standard Model from Muon Decay}

\affiliation{University of Alberta, Edmonton, AB, T6G 2J1, Canada}
\affiliation{University of British Columbia, Vancouver, BC, V6T 1Z1, Canada}
\affiliation{Kurchatov Institute, Moscow, 123182, Russia}
\affiliation{University of Montreal, Montreal, QC, H3C 3J7, Canada}
\affiliation{University of Regina, Regina, SK, S4S 0A2, Canada}
\affiliation{Texas A\&M University, College Station, TX 77843, U.S.A.}
\affiliation{TRIUMF, Vancouver, BC, V6T 2A3, Canada}
\affiliation{Valparaiso University, Valparaiso, IN 46383, U.S.A.}

\author{R.~Bayes}
\altaffiliation[Affiliated with: ]{Univ.\@ of Victoria,
Victoria, BC.}
\affiliation{TRIUMF, Vancouver, BC, V6T 2A3, Canada}

\author{J.F.~Bueno}
\affiliation{University of British Columbia, Vancouver, BC, V6T 1Z1, Canada}

\author{A.~Hillairet}
\altaffiliation[Affiliated with: ]{Univ.\@ of Victoria,
Victoria, BC.}
\affiliation{TRIUMF, Vancouver, BC, V6T 2A3, Canada}

\author{Yu.I.~Davydov}
\altaffiliation[Present Address: ]{JINR,
Dubna, Russia.}
\affiliation{TRIUMF, Vancouver, BC, V6T 2A3, Canada}

\author{P.~Depommier}
\affiliation{University of Montreal, Montreal, QC, H3C 3J7, Canada}

\author{W.~Faszer}
\affiliation{TRIUMF, Vancouver, BC, V6T 2A3, Canada}

\author{C.A.~Gagliardi}
\affiliation{Texas A\&M University, College Station, TX 77843, U.S.A.}

\author{A.~Gaponenko}
\altaffiliation[Present address: ]{LBNL, Berkeley, CA.}
\affiliation{University of Alberta, Edmonton, AB, T6G 2J1, Canada}

\author{D.R.~Gill}
\affiliation{TRIUMF, Vancouver, BC, V6T 2A3, Canada}

\author{A.~Grossheim}
\affiliation{TRIUMF, Vancouver, BC, V6T 2A3, Canada}

\author{P.~Gumplinger}
\affiliation{TRIUMF, Vancouver, BC, V6T 2A3, Canada}

\author{M.D.~Hasinoff}
\affiliation{University of British Columbia, Vancouver, BC, V6T 1Z1, Canada}

\author{R.S.~Henderson}
\affiliation{TRIUMF, Vancouver, BC, V6T 2A3, Canada}

\author{J.~Hu}
\altaffiliation[Present address: ]{AECL, Mississauga, ON, Canada}
\affiliation{TRIUMF, Vancouver, BC, V6T 2A3, Canada}

\author{D.D.~Koetke}
\affiliation{Valparaiso University, Valparaiso, IN 46383, U.S.A.}

\author{R.P.~MacDonald}
\affiliation{University of Alberta, Edmonton, AB, T6G 2J1, Canada}

\author{G.M.~Marshall}
\affiliation{TRIUMF, Vancouver, BC, V6T 2A3, Canada}

\author{E.L.~Mathie}
\affiliation{University of Regina, Regina, SK, S4S 0A2, Canada}

\author{R.E.~Mischke}
\email{mischke@triumf.ca}
\affiliation{TRIUMF, Vancouver, BC, V6T 2A3, Canada}

\author{K.~Olchanski}
\affiliation{TRIUMF, Vancouver, BC, V6T 2A3, Canada}

\author{A.~Olin}
\altaffiliation[Affiliated with: ]{Univ.\@ of Victoria,
Victoria, BC.}
\affiliation{TRIUMF, Vancouver, BC, V6T 2A3, Canada}

\author{R.~Openshaw}
\affiliation{TRIUMF, Vancouver, BC, V6T 2A3, Canada}

\author{J.-M.~Poutissou}
\affiliation{TRIUMF, Vancouver, BC, V6T 2A3, Canada}

\author{R.~Poutissou}
\affiliation{TRIUMF, Vancouver, BC, V6T 2A3, Canada}

\author{V.~Selivanov}
\affiliation{Kurchatov Institute, Moscow, 123182, Russia}

\author{G.~Sheffer}
\affiliation{TRIUMF, Vancouver, BC, V6T 2A3, Canada}

\author{B.~Shin}
\altaffiliation[Affiliated with: ]{Univ.\@ of Saskatchewan,
Saskatoon, SK.}
\affiliation{TRIUMF, Vancouver, BC, V6T 2A3, Canada}

\author{T.D.S.~Stanislaus}
\affiliation{Valparaiso University, Valparaiso, IN 46383, U.S.A.}

\author{R.~Tacik}
\affiliation{University of Regina, Regina, SK, S4S 0A2, Canada}

\author{R.E.~Tribble}
\affiliation{Texas A\&M University, College Station, TX 77843, U.S.A.}

\collaboration{TWIST Collaboration}
\noaffiliation

\date{\today}

\begin{abstract}
The TWIST Collaboration has completed a new measurement
of the energy-angle spectrum of positrons from the decay of highly polarized 
muons.  A simultaneous measurement of the muon decay parameters $\rho$, 
$\delta$, and $\Ppimu\xi$ tests the Standard Model (SM) in a purely leptonic 
process and provides improved limits for relevant extensions to the SM.
Specifically, for the generalized left-right symmetric model 
$|(g_R/g_L)\zeta|<0.020$ and $(g_L/g_R)m_2>$ 578 GeV$/c^2$, both 90\% C.L.
\end{abstract}


\maketitle
The maximal parity violation of the Standard Model (SM) charged weak 
interaction is empirically based.  Many natural SM extensions restore parity 
conservation at a higher mass scale with additional weak couplings.  Muon 
decay is an excellent laboratory to search for these couplings because the 
purely leptonic process can be calculated very precisely within the SM.  
This Letter presents a high precision measurement of the energy-angle 
spectrum of the positrons emitted in polarized muon decay, which provides 
new limits for the mass and mixing angle of the heavy $W$ in a 
class of left-right symmetric (LRS) models \cite{Herczeg}.

The most general Lorentz invariant, derivative-free muon decay matrix 
element \cite{Fetscher86} is described by 10 complex, model-independent 
couplings ($g^\gamma_{\epsilon\mu}$) involving left- and right-handed leptons 
($\epsilon,\mu$) in scalar, vector, and tensor interactions ($\gamma$). 
In the SM, $g^V_{LL}=1$, and the other nine constants are zero.
When only the positron energy and direction are measured, the muon decay 
spectrum can be described by four parameters
\cite{Michel50}, which are bilinear combinations of the 
$g^\gamma_{\epsilon\mu}$: the Michel parameter $\rho$, as well as
$\delta$, $\Pmu\xi$, and $\eta$. The differential decay rate is then
\begin{equation}
  \label{eq:michel}
  \nonumber
  \begin{split}
    \frac{d^2 \Gamma}{dx d(\cos \theta)} \propto &\, x^2 \bigg\{ (3 - 3x) + \frac{2}{3} \rho (4x-3) + 3\eta x_0\frac{(1-x)}{x} \\ & {} + \Pmu \xi \cos\theta \left[ (1-x) + \frac{2}{3} \delta (4x-3) \right] \bigg\},
  \end{split}
\end{equation}
with
\begin{displaymath}
x   =  \frac{E_e}{E_{max}},\
x_0 = \frac {m_e}{E_{max}},\
\Pmu  =  |\vec{{\cal P}}_{\mu}|,\ 
\cos{\theta}  =  \frac{\vec{{\cal P}}_{\mu} \cdot
  \vec{p}_e}{|\vec{{\cal P}}_{\mu}| \: |\vec{p}_e|}.
\end{displaymath}
The neutrino masses are neglected;
$E_{max}$ = 52.828 MeV. Radiative corrections \cite{Rad_corr} are not explicitly shown, 
but are significant and must be evaluated within the SM to a precision 
comparable to the experiment. The polarization of the muon from pion decay 
begins as $\Ppimu$ and may evolve over the $2.2\ \mu$s mean lifetime of the 
muon to become $\Pmu$ at the time of decay.
The SM predictions are $\rho = \delta = 3/4$, 
$\Ppimu = \xi = 1$, and $\eta = 0$.  
Precision measurements of these parameters test the SM predictions 
and are sensitive to extensions to the SM.

Prior to the TRIUMF Weak Interaction Symmetry Test (TWIST) experiment, 
$\rho$, $\delta$, and $\Ppimu\xi$ were known with uncertainties in the range 
of (3.5--8.5)$\times 10^{-3}$ \cite{Derenzo}. Intermediate TWIST results have 
already reduced those uncertainties to (0.7--3.8)$\times 10^{-3}$ 
\cite{MacDonald08,Jamieson06}. TWIST has now realized its goal 
of making about an order of magnitude improvement in each of the 
parameters. These final results of the 
experiment supersede those in our previous publications.

TWIST used highly polarized positive muons from
pions decaying on the surface of a graphite production target
irradiated with 500 MeV protons. The muons were transported by
the TRIUMF M13 surface muon channel to the entrance of a 2 T superconducting
solenoidal magnet and were guided by the 
field along its symmetry axis into the detector array
\cite{Henderson05}. A thin (239 $\mu$m) trigger scintillator identified muons
entering the detector. The muons were ranged to stop predominately
in a thin metal foil at the center of a symmetric stack of
high-precision, low-mass planar multiwire
chambers. There were 6 proportional and 22 drift chambers,
surrounded by helium gas, on each side of the stopping
target. Ionization of tracks from decay positrons was sampled by
the chambers and drift times were recorded by TDCs.

With a muon rate of (2--5)$\times 10^{3}$ s$^{-1}$, data sets of $10^9$
events could be obtained within a few days. Data sets were taken with two 
foil stopping targets, Ag (30.9 $\mu$m thick) and Al (71.6 $\mu$m thick), 
each $>$99.999\% pure. Sets were also taken with deliberately altered 
conditions to assist in studies of possible systematic errors.
A pair of time expansion chambers (TECs) \cite{Hu06} was inserted upstream 
of the solenoid to determine the incoming muon beam characteristics.
Because they caused multiple scattering and hence muon depolarization, 
the TECs were removed during most data taking. This phase of the 
experiment was completed in 2007.

Analyzed events were collected into two-dimensional (2D) distributions of 
positron angle and momentum (or energy) whose shape depends on the decay 
parameters. These distributions from data were compared to similar ones 
derived from a GEANT3 simulation \cite{Brun}. Both were subjected to
essentially the same analysis, allowing biases and inefficiencies
to be included in an equivalent way to reduce the dependence of
the result on the specific analysis procedure.  This places great
importance on the accuracy and detail of the simulation, which
includes not only standard physics processes but also a detailed
description of the beam, magnetic field, geometry, and detector
response. Decay parameters were extracted by fitting the
2D data distribution to that of a base distribution of 
simulated events, plus simulated distributions corresponding to the
first derivatives of the spectrum with respect
to decay parameters (or combinations thereof), yielding fit coefficients
$\Delta\rho$, $\Delta\xi$, and $\Delta\xi\delta$. The decay parameters used 
in the generation of the simulation were hidden, so the analysis was 
``blind'' \cite{MacDonald08}.
\begin{figure}[htb]
\centering
  \includegraphics[width=\columnwidth]{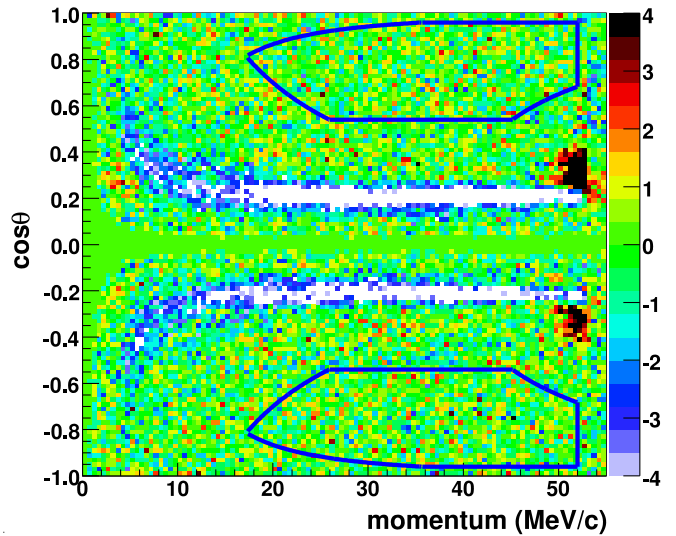}
  \caption{Residuals for the fit of one nominal data set to 
simulation in units of standard deviations. The fiducial region is outlined.}
  \label{fig:residuals}
\end{figure}

Fourteen data sets were used to extract $\rho$
and $\delta$, seven with each of the Ag and Al targets. Only nine
sets were used for $\Ppimu\xi$; the other five were 
acquired to test consistency and systematic effects with
altered beam, magnetic field, or muon multiple scattering, where the
depolarization was not optimally controlled. 
The residuals for the fit of one nominal data set in units of 
standard deviations ($\sigma$) are shown in 
Fig. \ref{fig:residuals}. A histogram of the residuals in the fiducial region, 
summed over all sets, has a mean of $-0.003 \pm 0.005$ and
$\sigma = 1.002 \pm 0.004$. Also shown is the range of 
$(p,\cos\theta)$ used to determine the decay parameters. 
The fiducial cuts are symmetric for upstream and downstream decays and were 
selected to maximize sensitivity to the decay parameters while reducing
systematic uncertainties. For all 14 data sets, there were
\scinot{11}{9} events, of which \scinot{0.55}{9} passed all event
selection criteria and were within the fiducial region. Simulation sets 
were typically 2.7 times larger than the corresponding data set.
The consistency of the data sets (statistical uncertainties only) was 
assessed by fits to constant means for 
the values of $\Delta\rho$, $\Delta\delta$, and $\Delta\Ppimu\xi$, which
gave reduced $\chi^2$ values of 14.0/13, 17.7/13, and 9.7/8, respectively.

The procedure of fitting the difference of two spectra in terms
of derivatives also provides a natural tool for the evaluation of systematic
uncertainties. The simulation was validated through comparison with observables
in the data 
that do not depend on muon decay parameter values, and the resulting 
uncertainties were factored into estimates of the systematic uncertainties. 
The sensitivities were obtained from the effect on the decay parameters when
an identified source of systematic uncertainty was changed (often
by an exaggerated amount) in one of the spectra. This was typically
achieved with two simulated spectra. 
The systematic uncertainties are listed, along with the
statistical errors, in Table \ref{tab:uncertainties}.
\begin{table}[ht]
\caption{Systematic uncertainties and statistical errors for $\rho$, $\delta$, 
and $\Ppimu\xi$.
\label{tab:uncertainties}}
\begin{center}
\begin{tabular}{|lc@{\extracolsep{0em}}cc|}
\hline
  Uncertainties
  & $\rho$
  & $\delta$
  & $\Ppimu\xi$ \\
  & ($\times 10^{-4}$) & ($\times 10^{-4}$) & ($\times 10^{-4}$) \\
\hline
\multicolumn{4}{|c|} {Target Independent Systematics} \\
\hline
Momentum calibration                 & 1.2  & 1.2  & 1.5 \\
Chamber response                     & 1.0  & 1.8  & 2.3 \\
Radiative corrections, $\eta$        & 1.3  & 0.6  & 1.2 \\
Resolution                           & 0.6  & 0.7  & 1.5 \\
Positron interactions\footnotemark[1]                & 0.5  & 0.2  & 0.4 \\
Others                               & 0.3  & 0.3  & 0.4 \\
\multicolumn{2}{|l}{Depolarization in fringe field}          &   & +15.8, -4.0 \\
\multicolumn{2}{|l}{Depolarization in stopping target}     &   & 3.2 \\
\multicolumn{2}{|l}{$\pi$\ decays in beamline}                      &   & 1.0 \\
\hline
\multicolumn{4}{|c|} {Uncertainties for Ag target}  \\
\hline
Bremsstrahlung rate                 & 1.8  & 1.6  & 0.5 \\
Ag thickness/stop position     & 3.8  & 6.4  & 0.6 \\
Statistical              & 1.2 & 2.1 & 4.2  \\
\hline
\multicolumn{4}{|c|}{Uncertainties for Al target} \\
\hline
Bremsstrahlung rate                 & 0.7  & 0.7  & 0.3 \\
Al thickness/stop position     & 0.2  & 0.8  & 0.8 \\
Statistical            & 1.4  & 2.4 & 3.9 \\
\hline
Weighted Systematic    & 2.3   & 2.7  & +16.5 -6.3\\
Weighted Statistical   & 1.2  & 2.1  & 2.9\\
\em{Total  Error}             & 2.6  & 3.4  & +16.8  -6.9 \\
\hline
\end{tabular}
\footnotetext[1]{excluding bremsstrahlung}
\end{center}
\end{table}

Notable improvements in the systematic uncertainties for $\rho$ and $\delta$
compared to our intermediate results \cite{MacDonald08} were achieved for 
positron interactions, chamber response, and momentum calibration.
The positron interactions systematic
addresses the possible inaccuracy in the simulation of reproducing
positron energy loss in the stopping target and detector
elements, primarily due to bremsstrahlung, delta-ray production,
and ionization. It was better constrained by comparisons of
identified interactions observed in the data and in the
simulation. Chamber response refers to the conversion of drift
times to spatial information used in track
fitting to evaluate the momentum and angle of each decay positron.
This was improved by more precise monitoring and control of
atmospheric influences that could change the chamber cell geometry.
In addition, a method was devised \cite{Grossheim10} to calibrate the 
chambers' space-time relations, for each plane, in both data and simulation, 
thereby reducing reconstruction biases. 
The maximum positron energy provides a calibration feature that
was used to reduce the energy scale uncertainty. Since energy loss
varies with the track angle linearly in $1\!
/\!(\cos \theta)$ due to the planar geometry of the detector, the
region near the kinematic endpoint of 52.8~MeV/c was
matched for data and simulation for small bins of $\cos \theta$.
The data-simulation relative energy calibration procedure has
undergone improvements to become more robust to fitting conditions.

The asymmetry parameter $\xi$ is also subject to uncertainties
from these sources, but they are overshadowed by
uncertainties unique to depolarization, as shown in Table
\ref{tab:uncertainties}. Depolarization in the fringe field and
in the muon stopping target result in $\Pmu < \Ppimu$ and constitute
the largest contributions to systematic uncertainties for
$\Ppimu\xi$. These uncertainties were improved considerably for this
analysis compared to the intermediate result \cite{Jamieson06}.
Improvements in the beam steering reduced the fringe field depolarization 
for a nominal data set to only \scinot{2.5}{-3}. The uncertainty 
depends on the accuracy of simulating the muon spin evolution as the beam
passes through significant radial field components at the
solenoid entrance. The essential 
ingredients are an accurate field map and precise knowledge of the
position and direction of the muons, as provided by the TECs.
Depolarization in the stopping target from muon spin relaxation
is assessed from the measured time dependence of the asymmetry.

After revealing the hidden parameters, the results for the three decay
parameters are consistent with the SM predictions. While the generalized 
matrix element treatment of Ref. \cite{Fetscher86} does not 
constrain the sign of deviations from the SM values for $\rho$, $\delta$, 
and $\xi$, the product $\Ppimu\xi\delta/\rho$ is constrained to be $\le 1.0$
and is 1.0 in the SM. This quantity defines the asymmetry 
between $\cos\theta = \pm 1$ at the maximum decay positron energy.
Our decay parameter values combined to give 
$\Ppimu\xi\delta/\rho = 1.00192 \ ^{+0.00167}_{-0.00066}$ (the errors account
for significant correlations), and the initial evaluation of 
$\Ppimu\xi\delta/\rho$ showed that the value for the Ag data was higher than 
that for Al by 3.8 $\sigma$. This apparent contradiction initiated 
an exhaustive reconsideration of effects that might have been 
overlooked in the blind analysis. The review showed that effects such as 
$\mu^+ \rightarrow e^+ X^0$\ decays (where $X^0$\ is a long-lived unobserved 
particle), an incorrect value of the $\eta$\ parameter,
or plausible errors in the radiative correction implementation were not 
responsible for the unexpected $\Ppimu\xi\delta/\rho$\ value. 

While no obvious mistake was uncovered in the estimates of the systematic 
uncertainties previously considered, we found that two corrections had been 
missed. A small correction was added for muon radiative decay 
($<$\scinot{1}{-4} for the Ag data and negligible for Al). Another 
correction was made for each set to account for a difference between the mean
muon stopping position for data and simulation.
We also concluded that the uncertainties for the two targets were sufficiently
different to merit dividing the systematic uncertainties into common and
target-dependent categories. The target independent systematics are unchanged 
from the blind analysis. Separate uncertainties for bremsstrahlung were
computed, and an additional sensitivity to the muon stopping position in the 
target was added.
\begin{figure}[htb]
  \includegraphics[width=\columnwidth]{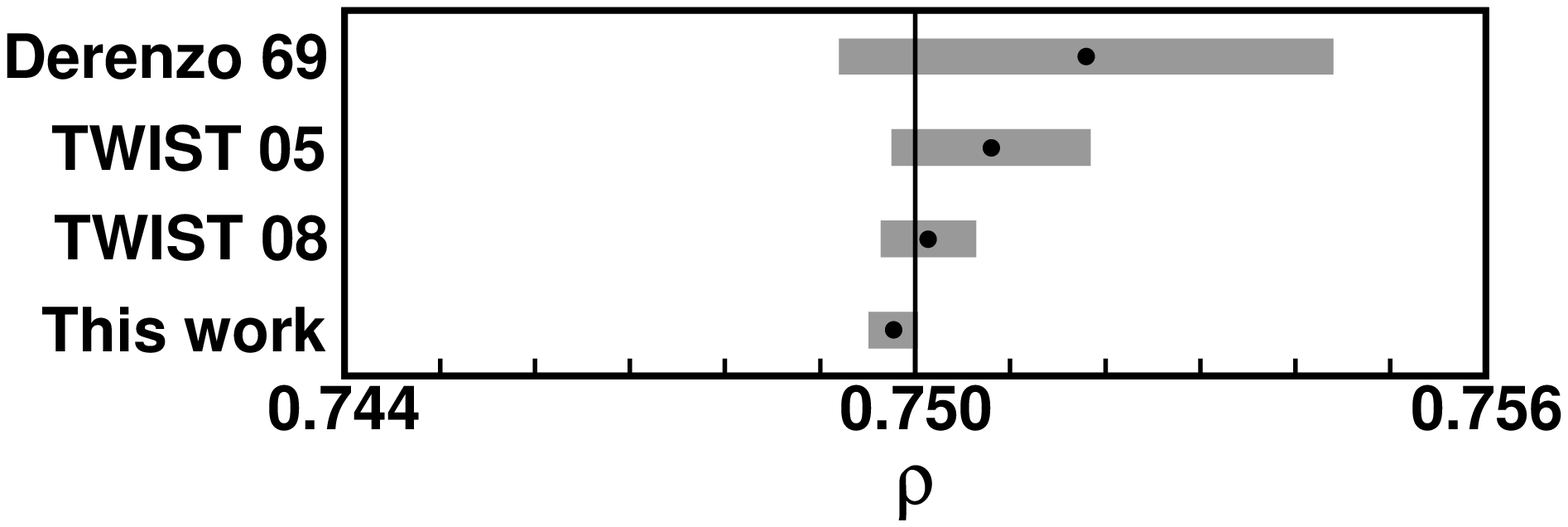}\\
  \includegraphics[width=\columnwidth]{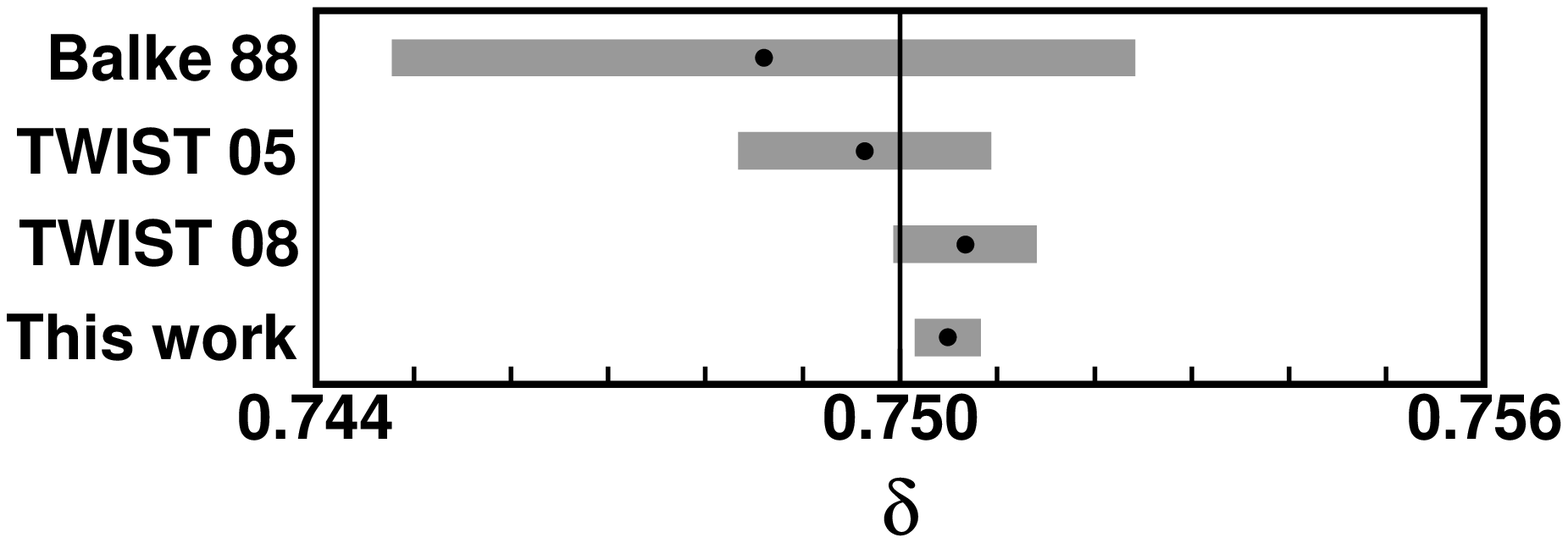}\\
  \includegraphics[width=\columnwidth]{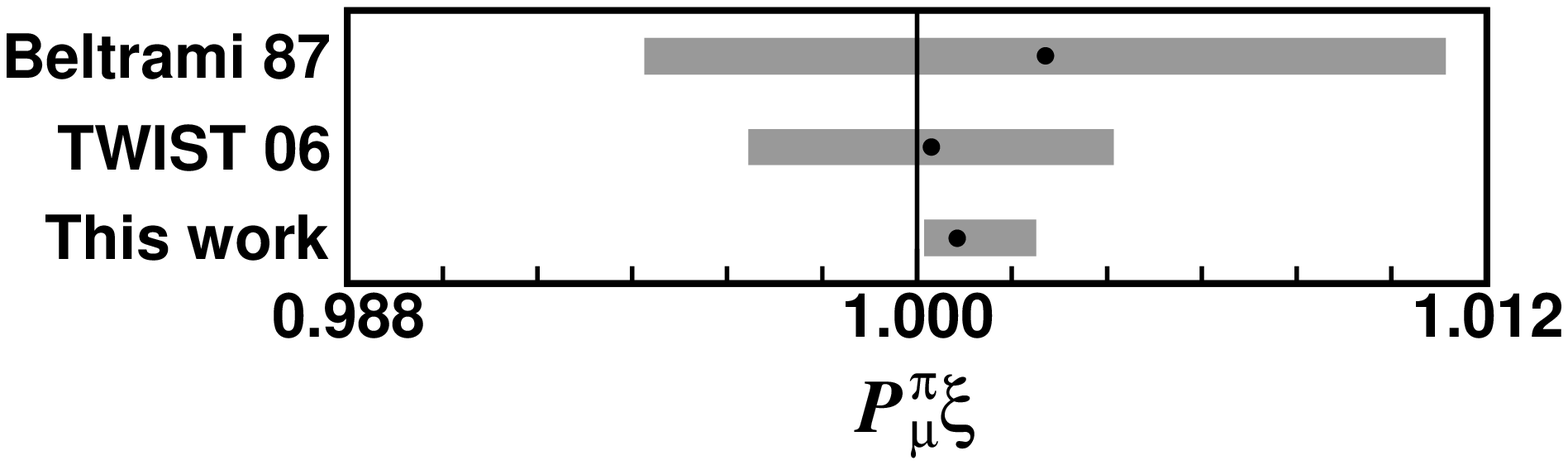}\\
  \caption{Summary of published central values and total uncertainties 
    for $\rho$, $\delta$, and $\Ppimu\xi$ 
    \cite{MacDonald08,Jamieson06,Derenzo,Musser,Gaponenko}, along with the 
    results of this analysis. Vertical lines represent the SM values.}
  \label{fig:bargraph_decaypars}
\end{figure}
 
With these changes the central values of $\rho$ and $\delta$ decreased from 
the blind results by 0.00014 and 0.00023, respectively.  $\Ppimu\xi$ is 
unchanged and its error reduced after including information from the 
measurement of $\delta$ in the five sets not used for $\Ppimu\xi$.
All uncertainties changed by $<$ 0.00006. The difference 
between targets for $\Ppimu\xi\delta/\rho$ is reduced to $\sim$1 $\sigma$, and
$\Ppimu\xi\delta/\rho = 1.00179 \ ^{+0.00156}_{-0.00071}$.
The revised results are compared to 
prior results in Fig. \ref{fig:bargraph_decaypars}. The values, including 
the uncertainties from Table 1, are:
\begin{eqnarray*}
\rho & = & 0.74977 \pm 0.00012(\mathrm{stat}) \pm 0.00023(\mathrm{syst}); \\
\delta & = & 0.75049 \pm 0.00021(\mathrm{stat}) \pm 0.00027(\mathrm{syst}); \\
\Ppimu\xi & = & 1.00084 \pm 0.00029(\mathrm{stat}) \ ^{+0.00165}_{-0.00063} (\mathrm{syst}).
\end{eqnarray*}

The decay parameters measured by TWIST contribute to a larger set
derived from other muon decay observables that can be analyzed in terms of 
the weak couplings $g^{\gamma}_{\epsilon\mu}$. A global analysis 
\cite{MacDonald08,Gagliardi05} imposes $\Ppimu\xi\delta/\rho \le 1.0$ 
and yields $\Ppimu\xi\delta/\rho>0.99909$ 
(90\% C.L.), compared to the pre-TWIST lower limit 
$\Ppimu\xi\delta/\rho>0.99682$ (90\% C.L.) \cite{Jodidio}. The global analysis
confirms consistency with the SM, where $g^V_{LL}$ is the only non-zero term. 
The TWIST results restrict the upper limits of other terms.
For example, the limit on the total right-handed muon coupling
\begin{eqnarray}
\nonumber
Q^{\mu}_R = \frac{1}{4}|g^S_{LR}|^2+\frac{1}{4}|g^S_{RR}|^2+|g^V_{LR}|^2+|g^V_{RR}|^2+3|g^T_{LR}|^2
\end{eqnarray}
is reduced by a factor of six from the pre-TWIST value to $<$\scinot{8.2}{-4} (90\% C.L.).
\begin{figure}[t]
\centering
  \includegraphics[width=\columnwidth]{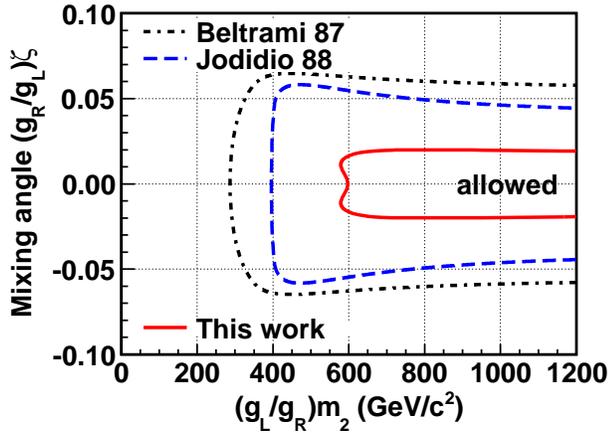}
  \caption{Allowed region (90\% C.L.) of mixing angle ($\zeta$) and heavy $W$ mass ($m_2$) 
for the general LRS model.}
  \label{fig:LRS}
\end{figure}

Left-right symmetric models extend the SM with a right-handed $W$ 
\cite{Herczeg}. In the generalized (or non-manifest) model no assumptions 
are made about the ratio of right- to left-handed couplings ($g_R/g_L$) or the 
form of the right-handed CKM matrix. In this case,
the TWIST result for $\rho$ provides the best limit on the mixing 
angle between the light and heavy mass eigenstates, $W_1$ and $W_2$.  
Our limit is $|(g_R/g_L)\zeta| < 0.020$ (90\% C.L.), compared to the 
pre-TWIST limit of $|(g_R/g_L)\zeta| < 0.066$.  The lower limit on the mass
of $W_2$ ($(g_L/g_R)m_2$) has been increased from 400 GeV$/c^2$ to 
578 GeV$/c^2$.  Coupled 
constraints on the mass for $(g_L/g_R)m_2$ and the mixing angle are shown in 
Fig. \ref{fig:LRS} where our limits are derived from a correlated 2D 
probability distribution from our measurements. These improved constraints 
will significantly impact predictions from the class of LRS models where 
the neutrinos are light compared to the muon mass. 

We thank all early TWIST collaborators and students
for their substantial contributions, as well as C. Ballard,
M. Goyette, and the TRIUMF cyclotron operations, beamlines, and
support personnel.
Computing resources were provided by WestGrid and Compute/Calcul Canada.
This work was supported in part by the Natural Sciences and Engineering 
Research Council and the National Research Council of Canada, the Russian 
Ministry of Science, and the U.S.\@ Department of Energy.

\end{document}